\documentclass[12pt]{article}

\usepackage{amsmath}
\usepackage{amssymb}

\usepackage{graphicx}

\begin{document}
\begin{flushright}
\hfill{ USTC-ICTS-08-20}\\
\end{flushright}
\vspace{4mm}

\begin{center}

{\Large \bf Statefinder Diagnosis for Ricci Dark Energy}

\vspace{8mm}

{\large Chao-Jun Feng}

\vspace{5mm}

{\em
 Institute of Theoretical Physics, CAS,\\
 Beijing 100080, P.R.China\\
 Interdisciplinary Center for Theoretical Study, USTC,\\
 Hefei, Anhui 230026, P.R.China
 }\\
\bigskip
fengcj@itp.ac.cn
\end{center}

\vspace{7mm}

\noindent Statefinder diagnostic is a useful method which can differ one dark energy model from each others.  In this
letter, we apply this method to a holographic dark energy model from Ricci scalar curvature, called the Ricci dark
energy model(RDE). We plot the evolutionary trajectories of this model in the statefinder parameter-planes, and it is
found that the parameter of this model plays a significant role from the statefinder viewpoint.  In a very special case
, the statefinder diagnostic fails to discriminate LCDM and RDE models, thus we apply a new diagnostic called the Om
diagnostic proposed recently to this model in this case in Appendix A and it works well.

\newpage
\section{Introduction}
The accelerating cosmic expansion first inferred from the observations of distant type Ia supernovae
\cite{Riess:1998cb} has strongly confirmed by some other independent observations, such as the cosmic microwave
background radiation (CMBR) \cite{Spergel:2006hy} and Sloan Digital Sky Survey (SDSS) \cite{:2007wu}. An exotic form of
negative pressure matter called dark energy is used to explain this acceleration. The simplest candidate of dark energy
is the cosmological constant $\Lambda$, whose energy density remains constant with time $\rho_\Lambda = \Lambda / {8\pi
G}$ and whose equation of motion is also fixed, $w_{\Lambda} = P_\Lambda / \rho_\Lambda = -1$ ($P_\Lambda$ is the
pressure) during the evolution of the universe. The cosmological model that consists of a mixture of the cosmological
constant and cold dark matter is called LCDM model, which provides an excellent explanation for the acceleration of the
universe phenomenon and other existing observational data. However, as is well know, this model faces two difficulties,
namely, the 'fine-tuning' problem and the 'cosmic coincidence' problem. The former also states: Why the cosmological
constant observed today is so much smaller than the Plank scale, while the latter states: Since the energy densities of
dark energy and dark matter scale so differently during the expansion of the universe, why they are at the same order
today? To alleviate or even solve these two problems, many dynamic dark energy models were proposed such as the
quintessence model rely on a scalar field minimally interacting with Einstein gravity. Here 'dynamic' means the
equation of state of the dark energy is no longer a constant but slightly evolves with time. Despite considerable works
on understanding the dark energy have been done, the nature of dark energy and its cosmological origin are still
enigmatic at present.\\

On the other hand, the problem of discriminating different dark energy models is now emergent. In order to solve this
problem, a sensitive and robust diagnostic for dark energy is a must. The statefinder parameter pair $\{r, s\}$
introduced by Sahni et al.\cite{Sahni:2002fz} and Alam et al.\cite{Alam:2003sc} is proven to be useful tools for this
purpose. The statefinder probes the expansion dynamics of the universe through high derivatives of the scale factor
$\ddot a$ \& $\dddot a$ and is a natural next step beyond the Hubble parameter $H\equiv \dot a/a$ and the deceleration
parameter q which depends upon $\ddot a$. The statefinder pair $\{ r, s\}$ is defined as
\begin{equation}\label{statefinder pair defined}
    r\equiv\frac{\dddot a}{aH^3}, \quad s\equiv\frac{r-1}{3(q-1/2)} .
\end{equation}
The statefinder pair is a 'geometrical' diagnostic in the sense that it is constructed from a space-time metric
directly, and it is more universal than 'physical' variables which depends upon properties of physical fields
describing dark energy, because physical variables are, of course, model-dependent. Usually one can plot the
trajectories in the $r-s$ plane corresponding to different dark energy models to see the qualitatively different
behaviors of them. The spatially flat LCDM scenario corresponds to a fixed point $\{ r, s \} = \{1,0\} $ in this
diagram. Departure of a given dark energy model from this fixed point provides a good way of establishing the
"distance" of this model from LCDM. As demonstrated in refs.\cite{Sahni:2002fz, Alam:2003sc, Zimdahl:2003wg} the
statefinder can successfully differentiate between a wide variety of dark energy models including the cosmological
constant, quintessence, the Chaplygin gas, braneworld
models and interacting dark energy models.\\

One can plot the current locations of the parameters $r$ and $s$ corresponding to different models in statefinder
parameter diagrams by theoretical calculating in these models. And on the other hand it can also be extracted from
experiment data such as the SNAP( SuperNovae Acceleration Probe ) data, with which combined the statefinder parameters
can serve as a versatile and powerful diagnostic of dark energy in the future. In this letter, we apply the statefinder
diagnostic to the Ricci dark energy model(RDE). It is found that the evolution behavior of the statefinder parameters
in this model is much like that in quiessence models, but in a very special case the statefinder diagnostic fails. In
Section II, we will briefly review RDE model, and apply the diagnostic to it in Section III. In the last section we
will give some conclusions. In the case of that the statefinder diagnostic fails, we apply a new diagnostic called the
Om diagnostic proposed recently to RDE model in Appendix A.

\section{Briefly Review on RDE}

Holographic principle \cite{Bousso:2002ju} regards black holes as the maximally entropic objects of a given region and
postulates that the maximum entropy inside this region behaves non-extensively, growing only as its surface area. Hence
the number of independent degrees of freedom is bounded by the surface area in Planck units, so an effective field
theory with UV cutoff $\Lambda$ in a box of size $L$ is not self consistent, if it does not satisfy the Bekenstein
entropy bound \cite{Bekenstein:1973ur} $ (L\Lambda)^3\leq S_{BH}=\pi L^2M_{pl}^2 $, where $M_{pl}^{-2}\equiv G $ is the
Planck mass and $S_{BH}$ is the entropy of a black hole of radius $L$ which acts as an IR cutoff. Cohen et.al.
\cite{Cohen:1998zx} suggested that the total energy in a region of size $L$ should not exceed the mass of a black hole
of the same size, namely $ L^3\Lambda^4\leq LM_p^2 $. Therefore the maximum entropy is $S^{3/4}_{BH}$. Under this
assumption, Li \cite{Li:2004rb} proposed the holographic dark energy as follows
\begin{equation}\label{li}
    \rho_\Lambda = 3c^2M_p^2 L^{-2}
\end{equation}
where $c^2$ is a dimensionless constant. Since the holographic dark energy with Hubble horizon as its IR cutoff does
not give an accelerating universe \cite{Hsu:2004ri}, Li suggested to use the future event horizon instead of Hubble
horizon and particle horizon, then this model gives an accelerating universe and is consistent with current
observation\cite{Li:2004rb, Huang:2004ai}. For the recent works on holographic dark energy, see ref.
\cite{Zhang:2007sh,Sadjadi:2007ts, Saridakis:2007cy}. \\

Recently, Gao et.al \cite{Gao:2007ep} have proposed a holographic dark energy model in which the future event horizon
is replaced by the inverse of the Ricci scalar curvature, and they call this model the Ricci dark energy model(RDE).
This model does not only avoid the causality problem and is phenomenologically viable, but also solve the coincidence
problem of dark energy. The Ricci curvature of FRW universe is given by
\begin{equation}\label{Ricci}
    R = -6(\dot H + 2H^2 + \frac{k}{a^2}) \, ,
\end{equation}
where dot denotes a derivative with respect to time $t$ and $k$ is the spatial curvature. They introduced a holographic
dark energy proportional to the Ricci scalar
\begin{equation}\label{Ricci DE}
    \rho_X = \frac{3\alpha}{8\pi G} \left(\dot H + 2H^2 + \frac{k}{a^2}\right) \propto R
\end{equation}
where the dimensionless coefficient $\alpha$ will be determined by observations and they call this model the Ricci dark
energy model. Solving the Friedmann equation they find the result
\begin{equation}
   \frac{8\pi G }{3H^2_0}\rho_X  = \frac{\alpha}{2-\alpha}\Omega_{m0}e^{-3x} + f_0e^{-(4-\frac{2}{\alpha})x}
\end{equation}
where $\Omega_{m0} \equiv 8\pi G\rho_{m0}/3H^2_0$, $x = \ln{a}$ and $f_0$ is an integration constant. Substituting the
expression of $\rho_X$ into the conservation equation of energy,
\begin{equation}\label{conservation law}
    p_X = -\rho_X-\frac{1}{3}\frac{d\rho_X}{dx}
\end{equation}
we get the pressure of dark energy
\begin{equation}\label{pressure of X}
    p_X = -\frac{3H^2_0}{8\pi G }\left(\frac{2}{3\alpha}-\frac{1}{3}\right)f_0e^{-(4-\frac{2}{\alpha})x}
\end{equation}
Taking the observation values of parameters they find the $\alpha \simeq 0.46 $ and $f_0 \simeq 0.65$
\cite{Gao:2007ep}. The evolution of the equation of state $w_X \equiv p_X / \rho_X $ of dark energy is the following.
At high redshifts the value of $w_X $ is closed to zero, namely the dark energy behaves like the cold dark matter, and
nowadays $w_X $ approaches $-1$ as required and in the future the dark energy will be phantom. The energy density of
RDE during big bang nucleosynthesis(BBN) is so much smaller than that of other components of the universe ($\Omega_X
|_{1MeV}<10^{-6}\ll 0.1$ when $\alpha<1$), so it does not affect BBN procedure. Further more this model can avoid the
age problem and the causality problem. In next section we will study RDE model from the statefinder diagnostic
viewpoint.

\section{Statefinder Diagnostic for RDE}
The statefinder parameters can be expressed in terms of the total energy density $\rho$ and the total pressure $p$ in
the universe as follows
\begin{equation}\label{statefinder r s in p}
    r = 1 + \frac{9(\rho+p)}{2\rho}\frac{\dot p}{\dot\rho} \,, \quad s = \frac{(\rho+p)}{p}\frac{\dot p}{\dot\rho}
    \, .
\end{equation}
The deceleration parameter $q \equiv -\ddot a/(aH^2)$ can be also expressed in terms of $\rho$ and $p$
\begin{equation}\label{deceleration}
    q=\frac{1}{2}\left( 1 + \frac{3p}{\rho} \right)
\end{equation}
Assuming the universe is well described by a two component fluid consisting of non-relativistic matter (CDM+baryons)
with negligible pressure, i.e. $p_m << \rho_m$ and dark energy, namely, $\rho = \rho_m + \rho_X$, and $p \approx p_X$,
we obtain the statefinder parameters for RDE model as follows
\begin{eqnarray}\label{rs for RDE}
  \nonumber 
  r &=& 1 - \left(\frac{1}{\alpha^2}\right)\frac{\left(2-\alpha\right)\left(2\alpha-1\right)f_0e^{-(4-\frac{2}{\alpha})x}}{\frac{2}{2-\alpha}\Omega_{m0}e^{-3x} + f_0e^{-(4-\frac{2}{\alpha})x} }\, , \\
  s &=& \frac{2}{3}\left( 2 - \frac{1}{\alpha} \right),
\end{eqnarray}
and the deceleration parameter
\begin{equation}\label{q for RDE}
    q = \frac{1}{2}\left( 1 - \left(\frac{1}{\alpha}\right)\frac{\left(2-\alpha\right)f_0e^{-(4-\frac{2}{\alpha})x}}{\frac{2}{2-\alpha}\Omega_{m0}e^{-3x} + f_0e^{-(4-\frac{2}{\alpha})x}} \right).
\end{equation}
From eq.(\ref{rs for RDE}), one can see that $s = 0$, $r=1$ if $\alpha = 0.5$ and no matter what value $f_0$ is, and
this point in the $r-s$ plane is the very fixed point corresponding to LCDM model. Thus, the statefinder diagnostic
fails to discriminate between the LCDM and RDE model in this case. If $\alpha < 0.5$, then the trajectory will lying in
the region $r > 1$, $s < 0$.\\

As an example, we plot the statefinder diagrams in the $r-s$ plane and $r-q$ plane as a complementarity with $\alpha =
0.46$ and $f_0 = 0.65$ obtained in ref.\cite{Gao:2007ep} in Fig.1 and Fig.2.\\

\bigskip{
    \vbox{
            {
                \nobreak
                \centerline
                {
                    \includegraphics[scale=1.0]{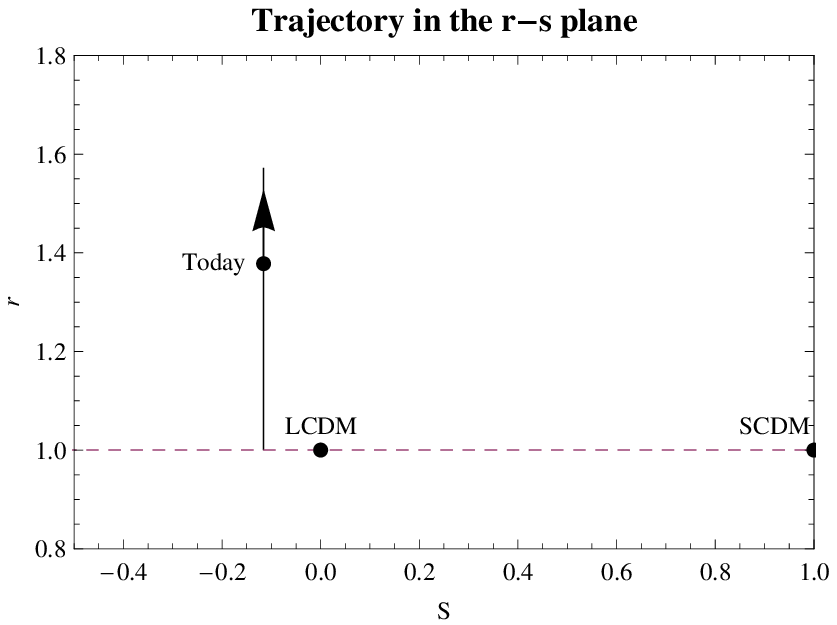}
                }
                \nobreak
                \bigskip
                {\raggedright\it \vbox
                    {
                        {\bf Figure 1.}
                        {\it Evolution trajectory in the statefinder $r-s$ plane for RDE with $\alpha = 0.46$ and
                        $f_0=0.65$.
                        }
                    }
                }

            }
        }
\bigskip}
In Fig.1, LCDM scenario corresponds to a fixed point $s=0, r=1$, and the SCDM (standard cold matter) scenario
corresponds to the point $s=1,r=1$. For RDE model, the trajectory is a vertical segment, i.e. s is a constant during
the evolution of the universe, while r monotonically increases from $1$ to $1- (2-\alpha)(2\alpha-1)/\alpha^2 \approx
1.58$. The location of today's point is $s=-0.12,r=1.38$, thus the 'distance' from RDE model to  LCDM model can be
easily identified in this diagram. The trajectories for the so-called 'quiessence' model ($w$ is a constant) are also
vertical segments, but in that model, $r$ decreases monotonically from $1$ to $1+9w(1+w)/2$ while $s$ remains constant
at $1+w$ \cite{Sahni:2002fz, Alam:2003sc}.\\

In fact, the statefinder diagnostic can also discriminate between other dark energy models effectively. For example,
the trajectories for the Chaplygin gas and the quintessence(inverse power low) models are similar to arcs of a parabola
(downward and upward) lying in the regions $s <0, r>1$ and $s>0, r<1$ respectively. For holographic dark energy model
with the future event horizon as IR cutoff, commences its evolution from the point $s=2/3, r=1$, through an arc
segment, and ends it at LCDM fixed point in the future\cite{Zimdahl:2003wg}. Therefore, the distinctive trajectories
corresponding to various dark energy scenario in the $r-s$ plane demonstrate quite strikingly the contrasting behaviors
of dark energy
models.\\

\bigskip{
    \vbox{
            {
                \nobreak
                \centerline
                {
                    \includegraphics[scale=1.0]{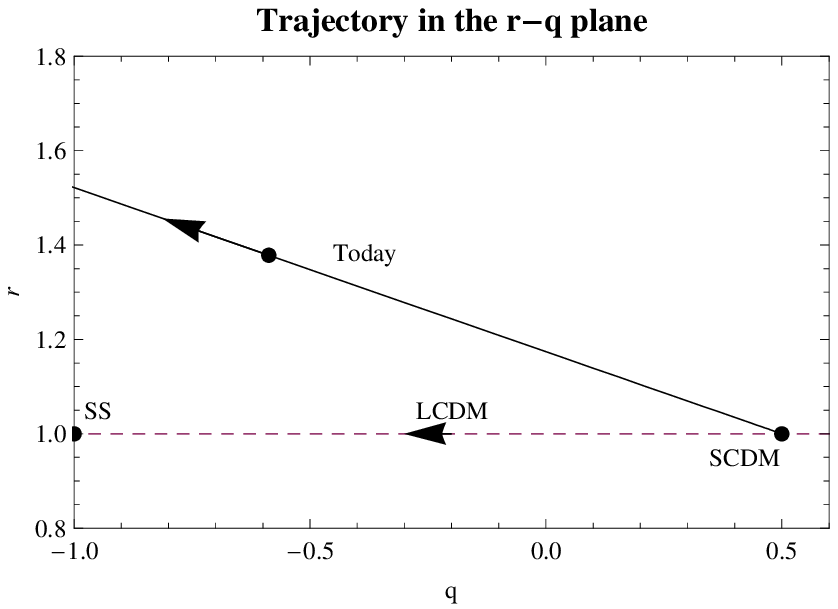}
                }
                \nobreak
                \bigskip
                {\raggedright\it \vbox
                    {
                        {\bf Figure 2.}
                        {\it Evolution trajectory in the statefinder $r-q$ plane for RDE with $\alpha = 0.46$ and
                        $f_0=0.65$. The solid line represents the RDE model, and the dashed line the LCDM as comparison.
                        The location of today's point is $(-0.59, 1.38)$.
                        }
                    }
                }

            }
        }
\bigskip}
In Fig.2, we clearly see that both LCDM scenario and RDE model commence evolving from the same point in the past
$q=0.5,r=1$, which corresponds to a matter dominated SCDM universe. However, in LCDM model the trajectory will end
their evolution at the point $q=-1,r=1$ which corresponds to a steady state cosmology(SS), i.e. the de Sitter
expansion, while that in RDE model does not. In ref.\cite{Zimdahl:2003wg}, the trajectory in holographic dark energy
model with the future event horizon (HDE) has the same starting point and the same ending point as that in LCDM model
. Thus, RDE model is also different from HDE model from the statefinder viewpoint.\\

If $\alpha > 0.5$ , the sign of $s$ becomes positive and $r<1$ if $\alpha$ is also smaller than $2$, but the value of
$r$ will lager than $1$ if $\alpha > 2$, see eq.(\ref{rs for RDE}). Thus, the determining of the value of $\alpha$ is a
key point to the feature of RDE model and we hope the future high precision experiments may provide sufficiently large
amount of precise data to be capable of determining the value of $\alpha$.

\section{Conclusions}
In this letter, we have apply the statefinder diagnostic to the Ricci dark energy model, and plot the trajectories in
the $r-s$ and $r-q$ planes in the case of $\alpha=0.46$ and $f_0=0.65$. Here we have used the values of $\alpha$ and
$f_0$ that founded in ref.\cite{Gao:2007ep}, but other values are also possible. Different values of $\alpha$ will
determine different evolutions of the statefinder parameters, so the determining of $\alpha$ from more precise data
provided by future experiments will be needed. In a very special case that $\alpha = 0.5$, the statefinder pair $\{r,
s\}$ fails to discriminate LCDM model and RDE model, because they give the same fixed point $r=1, s=0$ in the $r-s$
diagram. The difference of these two models is the evolution of the equation of state $w$, which is a constant that
equals $-1$ in the former and a time-dependent variable in the latter. Recently, a new diagnostics called the $Om$
diagnostic of dark energy is proposed in ref.\cite{Sahni:2008xx}. In Appendix A, we will apply this diagnostic to RDE
model in the case of $\alpha=0.5$ in order to differentiate LCDM model and RDE model. The result indicates that this
new diagnostic really works well for this purpose.

\section*{Appendix A}
The definition of the $Om$ diagnostic is \cite{Sahni:2008xx}
\begin{equation}\label{om diag}
    Om(x)\equiv \frac{h^2(x)-1}{e^{-3x}-1} \, ,
\end{equation}
where $h(x)\equiv H(x)/H_0$. For dark energy with a constant dark energy equation of state $w = const.$ in the
spatially flat universe,
\begin{equation}\label{h const}
    h^2(x) = \Omega_{m0}e^{-3x}+(1-\Omega_{m0})e^{-3(1+w)x}.
\end{equation}
Consequently,
\begin{equation}\label{om for DE}
    Om(x) = \Omega_{m0}+(1-\Omega_{m0}) \frac{e^{-3(1+w)x}-1}{e^{-3x}-1} \, ,
\end{equation}
from where we get
\begin{equation}\label{om for cc}
    Om(x) = \Omega_{m0}
\end{equation}
in the LCDM model. Authors in ref.\cite{Sahni:2008xx} conclude that: $Om(x)-\Omega_{m0} = 0$, if and only if dark
energy is a
cosmological constant. \\

From eq.(8) in \cite{Gao:2007ep}, we get
\begin{equation}\label{h for RDE}
    h^2(x) = \frac{2}{2-\alpha}\Omega_{m0}e^{-3x} + f_0e^{-(4-\frac{2}{\alpha})x}
\end{equation}
in RDE model, then the $Om$ diagnostic for it is
\begin{equation}\label{om for rde}
    Om(x) = \frac{\frac{2}{2-\alpha}\Omega_{m0}e^{-3x} + f_0e^{-(4-\frac{2}{\alpha})x}-1}{e^{-3x}-1} \,.
\end{equation}
As an example, we take $\alpha=0.5$ and $\Omega_{m0}=0.27$ to plot the evolutions of $Om(x)$ corresponding to $f_0=0.5$
, $ 0.65$ and $0.8$ in Fig.3.\\

\bigskip{
    \vbox{
            {
                \nobreak
                \centerline
                {
                    \includegraphics[scale=1.0]{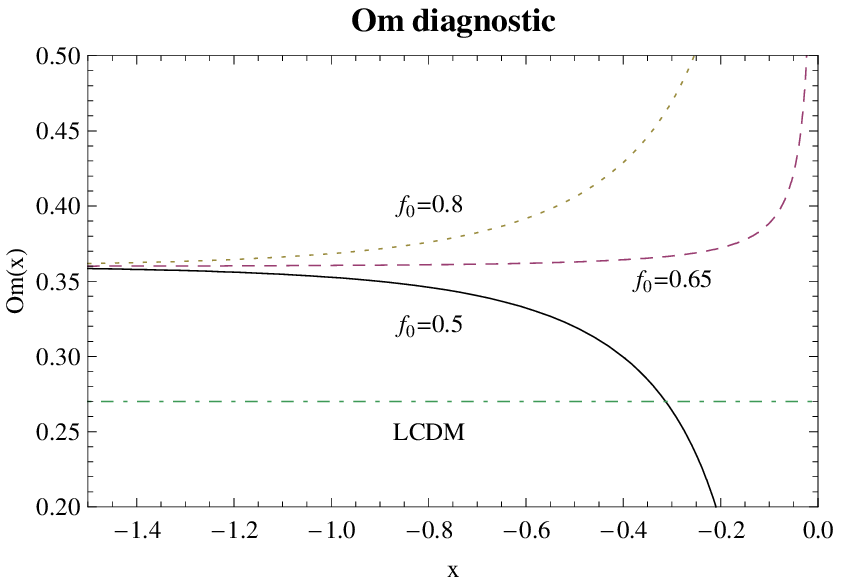}
                }
                \nobreak
                \bigskip
                {\raggedright\it \vbox
                    {
                        {\bf Figure 3.}
                        {\it The Om diagnostic for RDE with $\alpha=0.5$ , $\Omega_{m0}=0.27$ as well as $f_0=0.5$ ,$ 0.65$ and $0.8$, respectively.
                        }
                    }
                }

            }
        }
\bigskip}
Thus, one can easily find the difference between the LCDM model and RDE model from Fig.3. Especially the difference is
much larger near present, i.e. $x\sim 0$. Here $f_0$ plays a important role to determine the evolution
of $Om$ , so $f_0$' value is also hoped to be determined from future precise data as $\alpha$. \\

\section*{ACKNOWLEDGEMENTS}
The author would like to thank Miao Li for a careful reading of the manuscript and valuable suggestions. We are
grateful to Tower Wang for useful discussions.


\newpage
\begin{thebibliography}{99}
\bibitem{Riess:1998cb}
  A.~G.~Riess {\it et al.}  [Supernova Search Team Collaboration],
   ``Observational Evidence from Supernovae for an Accelerating Universe and a
  Astron.\ J.\  {\bf 116}, 1009 (1998)
  [arXiv:astro-ph/9805201].

  S.~Perlmutter {\it et al.}  [Supernova Cosmology Project Collaboration],
  Astrophys.\ J.\  {\bf 517}, 565 (1999)
  [arXiv:astro-ph/9812133].

\bibitem{Spergel:2006hy}
  D.~N.~Spergel {\it et al.}  [WMAP Collaboration],
   ``Wilkinson Microwave Anisotropy Probe (WMAP) three year results:
  Astrophys.\ J.\ Suppl.\  {\bf 170}, 377 (2007)
  [arXiv:astro-ph/0603449].

  E.~Komatsu {\it et al.}  [WMAP Collaboration],
  arXiv:0803.0547 [astro-ph].

\bibitem{:2007wu}
  J.~K.~Adelman-McCarthy {\it et al.}  [SDSS Collaboration],
  arXiv:0707.3413 [astro-ph].

\bibitem{Sahni:2002fz}
  V.~Sahni, T.~D.~Saini, A.~A.~Starobinsky and U.~Alam,
  JETP Lett.\  {\bf 77}, 201 (2003)
  [Pisma Zh.\ Eksp.\ Teor.\ Fiz.\  {\bf 77}, 249 (2003)]
  [arXiv:astro-ph/0201498].

\bibitem{Alam:2003sc}
  U.~Alam, V.~Sahni, T.~D.~Saini and A.~A.~Starobinsky,
  Mon.\ Not.\ Roy.\ Astron.\ Soc.\  {\bf 344}, 1057 (2003)
  [arXiv:astro-ph/0303009].

\bibitem{Zimdahl:2003wg}
  W.~Zimdahl and D.~Pavon,
  Gen.\ Rel.\ Grav.\  {\bf 36}, 1483 (2004)
  [arXiv:gr-qc/0311067].

  X.~Zhang,
  Phys.\ Lett.\  B {\bf 611}, 1 (2005)
  [arXiv:astro-ph/0503075].

  X.~Zhang,
  Int.\ J.\ Mod.\ Phys.\  D {\bf 14}, 1597 (2005)
  [arXiv:astro-ph/0504586].

  M.~P.~Dabrowski,
  Phys.\ Lett.\  B {\bf 625}, 184 (2005)
  [arXiv:gr-qc/0505069].

  P.~x.~Wu and H.~w.~Yu,
  Int.\ J.\ Mod.\ Phys.\  D {\bf 14}, 1873 (2005)
  [arXiv:gr-qc/0509036].

  M.~G.~Hu and X.~H.~Meng,
  Phys.\ Lett.\  B {\bf 635}, 186 (2006)
  [arXiv:astro-ph/0511615].

  E.~E.~O.~Ishida,
  Braz.\ J.\ Phys.\  {\bf 35}, 1172 (2005)
  [arXiv:astro-ph/0609614].

  M.~R.~Setare, J.~Zhang and X.~Zhang,
  JCAP {\bf 0703}, 007 (2007)
  [arXiv:gr-qc/0611084].

  B.~R.~Chang, H.~Y.~Liu, L.~X.~Xu, C.~W.~Zhang and Y.~L.~Ping,
  JCAP {\bf 0701}, 016 (2007)
  [arXiv:astro-ph/0612616].

  Y.~Shao and Y.~Gui,
  Mod.\ Phys.\ Lett.\  A {\bf 23}, 65 (2008)
  [arXiv:gr-qc/0703111].

  Z.~L.~Yi and T.~J.~Zhang,
  Phys.\ Rev.\  D {\bf 75}, 083515 (2007)
  [arXiv:astro-ph/0703630].

  B.~Chang, H.~Liu, L.~Xu and C.~Zhang,
  Mod.\ Phys.\ Lett.\  A {\bf 23}, 269 (2008)
  [arXiv:0704.3670 [astro-ph]].

  C.~Bao-rong, L.~Hong-ya, X.~Li-xin and Z.~Cheng-wu,
  arXiv:0704.3768 [astro-ph].

  J.~Zhang, X.~Zhang and H.~Liu,
  Phys.\ Lett.\  B {\bf 659}, 26 (2008)
  [arXiv:0705.4145 [astro-ph]].

  H.~Wei and R.~G.~Cai,
  Phys.\ Lett.\  B {\bf 655}, 1 (2007)
  [arXiv:0707.4526 [gr-qc]].

  W.~Zhao,
  arXiv:0711.2319 [gr-qc].

  D.~J.~Liu and W.~Z.~Liu,
  Phys.\ Rev.\  D {\bf 77}, 027301 (2008)
  [arXiv:0711.4854 [astro-ph]].

  G.~Panotopoulos,
  Nucl.\ Phys.\  B {\bf 796}, 66 (2008)
  [arXiv:0712.1177 [astro-ph]].

  Z.~G.~Huang, X.~M.~Song, H.~Q.~Lu and W.~Fang,
  Astrophys.\ Space Sci.\  {\bf 315}, 175 (2008)
  [arXiv:0802.2320 [hep-th]].

  Z.~G.~Huang and H.~Q.~Lu,
  arXiv:0802.2321 [hep-th].

  W.~Z.~Liu and D.~J.~Liu,
  arXiv:0803.4039 [astro-ph].




\bibitem{Bousso:2002ju}
  R.~Bousso,
  Rev.\ Mod.\ Phys.\  {\bf 74}, 825 (2002)
  [arXiv:hep-th/0203101].

\bibitem{Bekenstein:1973ur}
  J.~D.~Bekenstein,
  Phys.\ Rev.\  D {\bf 7}, 2333 (1973).

  J.~D.~Bekenstein,
   ``A Universal Upper Bound On The Entropy To Energy Ratio For Bounded
  Phys.\ Rev.\  D {\bf 23}, 287 (1981).

\bibitem{Cohen:1998zx}
  A.~G.~Cohen, D.~B.~Kaplan and A.~E.~Nelson,
  Phys.\ Rev.\ Lett.\  {\bf 82}, 4971 (1999)
  [arXiv:hep-th/9803132].

\bibitem{Li:2004rb}
  M.~Li,
  Phys.\ Lett.\  B {\bf 603}, 1 (2004)
  [arXiv:hep-th/0403127].

\bibitem{Hsu:2004ri}
  S.~D.~H.~Hsu,
  Phys.\ Lett.\  B {\bf 594}, 13 (2004)
  [arXiv:hep-th/0403052].

\bibitem{Huang:2004ai}
  Q.~G.~Huang and M.~Li,
  JCAP {\bf 0408}, 013 (2004)
  [arXiv:astro-ph/0404229].

  Q.~G.~Huang and M.~Li,
  JCAP {\bf 0503}, 001 (2005)
  [arXiv:hep-th/0410095].


\bibitem{Zhang:2007sh}
  X.~Zhang and F.~Q.~Wu,
  Phys.\ Rev.\  D {\bf 76}, 023502 (2007)
  [arXiv:astro-ph/0701405].

  Y.~S.~Myung,
  Phys.\ Lett.\  B {\bf 649}, 247 (2007)
  [arXiv:gr-qc/0702032].

  H.~b.~Zhang, W.~Zhong, Z.~H.~Zhu and S.~He,
  Phys.\ Rev.\  D {\bf 76}, 123508 (2007)
  [arXiv:0705.4409 astro-ph].

  Z.~Y.~Sun and Y.~G.~Shen,
  Int.\ J.\ Theor.\ Phys.\  {\bf 46}, 877 (2007).

  Y.~S.~Myung,
  Phys.\ Lett.\  B {\bf 652}, 223 (2007)
  [arXiv:0706.3757 gr-qc].

  C.~Feng, B.~Wang, Y.~Gong and R.~K.~Su,
  JCAP {\bf 0709}, 005 (2007)
  [arXiv:0706.4033 astro-ph].

  H.~Wei and S.~N.~Zhang,
  Phys.\ Rev.\  D {\bf 76}, 063003 (2007)
  [arXiv:0707.2129 astro-ph].

  B.~Guberina,
  [arXiv:0707.3778 gr-qc].

  B.~C.~Paul, P.~Thakur and A.~Saha,
  [arXiv:0707.4625 gr-qc].

  J.~f.~Zhang, X.~Zhang and H.~y.~Liu,
  Eur.\ Phys.\ J.\  C {\bf 52}, 693 (2007)
  [arXiv:0708.3121 hep-th].

  C.~J.~Feng,
  [arXiv:0709.2456 hep-th].

  Y.~Z.~Ma and Y.~Gong,
  [arXiv:0711.1641 astro-ph].

  R.~Horvat,
  [arXiv:0711.4013 gr-qc].

  E.~Elizalde, S.~Nojiri, S.~D.~Odintsov and P.~Wang,
  Phys.\ Rev.\  D {\bf 71}, 103504 (2005)
  [arXiv:hep-th/0502082].

  S.~Nojiri and S.~D.~Odintsov,
  Gen.\ Rel.\ Grav.\  {\bf 38}, 1285 (2006)
  [arXiv:hep-th/0506212].

\bibitem{Sadjadi:2007ts}
  H.~M.~Sadjadi,
  JCAP {\bf 0702}, 026 (2007)
  [arXiv:gr-qc/0701074].

  M.~R.~Setare,
  JCAP {\bf 0701}, 023 (2007)
  [arXiv:hep-th/0701242].

  M.~R.~Setare and E.~C.~Vagenas,
  [arXiv:0704.2070 hep-th].

  J.~Zhang, X.~Zhang and H.~Liu,
  Phys.\ Lett.\  B {\bf 659}, 26 (2008)
  [arXiv:0705.4145 astro-ph].

  Q.~Wu, Y.~Gong, A.~Wang and J.~S.~Alcaniz,
  Phys.\ Lett.\  B {\bf 659}, 34 (2008)
  [arXiv:0705.1006 astro-ph].

  K.~Y.~Kim, H.~W.~Lee and Y.~S.~Myung,
  Mod.\ Phys.\ Lett.\  A {\bf 22}, 2631 (2007)
  [arXiv:0706.2444 gr-qc].

  M.~R.~Setare,
  [arXiv:0708.3284 hep-th].

\bibitem{Saridakis:2007cy}
  E.~N.~Saridakis,
  Phys.\ Lett.\  B {\bf 660}, 138 (2008)
  [arXiv:0712.2228 hep-th].

  E.~N.~Saridakis,
  JCAP {\bf 0804}, 020 (2008)
  [arXiv:0712.2672 astro-ph].

  X.~Wu and Z.~H.~Zhu,
  Phys.\ Lett.\  B {\bf 660}, 293 (2008)
  [arXiv:0712.3603 astro-ph].

  E.~N.~Saridakis,
  Phys.\ Lett.\  B {\bf 661}, 335 (2008)
  [arXiv:0712.3806 [gr-qc].

  A.~A.~Sen and D.~Pavon,
  [arXiv:0801.0280 astro-ph].

  M.~Li, C.~Lin and Y.~Wang,
  [arXiv:0801.1407 astro-ph].

  K.~Karwan,
  JCAP {\bf 0805}, 011 (2008)
  [arXiv:0801.1755 astro-ph].

  M.~R.~Setare and E.~C.~Vagenas,
  [arXiv:0801.4478 hep-th].

  A.~J.~M.~Medved,
  [arXiv:0802.1753 hep-th].

  Y.~S.~Myung and M.~G.~Seo,
  [arXiv:0803.2913 gr-qc].

  B.~Nayak and L.~P.~Singh,
  arXiv:0803.2930 [gr-qc].

  L.~Xu and J.~Lu,
  arXiv:0804.2925 [astro-ph].

  K.~Y.~Kim, H.~W.~Lee and Y.~S.~Myung,
  arXiv:0805.3941 [gr-qc].

  C.~J.~Feng,
  arXiv:0806.0673 [hep-th].

  H.~Mohseni Sadjadi and N.~Vadood,
  arXiv:0806.2767 [gr-qc].

  R.~Horvat,
  arXiv:0806.4825 [gr-qc].

  N.~Cruz, S.~Lepe, F.~Pena and J.~Saavedra,
  arXiv:0807.3854 [gr-qc].

  Y.~Bisabr,
  arXiv:0808.1424 [gr-qc].



\bibitem{Gao:2007ep}
  C.~Gao, X.~Chen and Y.~G.~Shen,
  [arXiv:0712.1394 astro-ph].

\bibitem{Sahni:2008xx}
  V.~Sahni, A.~Shafieloo and A.~A.~Starobinsky,
  arXiv:0807.3548 [astro-ph].


\end{thebibliography}
\end{document}